\documentclass[prb,twocolumn,showpacs,preprintnumbers,amsmath,amssymb,floats]{revtex4-1}
\pdfoutput=1
\usepackage{graphicx}
\usepackage{amsmath}
\usepackage{epstopdf}
\usepackage{amsbsy}
\usepackage{color}
\usepackage{dcolumn}
\usepackage{bm}

\begin{document}

\title{Incommensurate charge ordered states in the $\mathit{t-t^{\prime}-J}$ model}
\author{Peayush Choubey$^1$, Wei-Lin Tu$^{2,3,4}$, Ting-Kuo Lee$^4$, and  P. J. Hirschfeld$^1$}
\affiliation{
$^1$Department of Physics, University of Florida, Gainesville, Florida 32611, USA \\
$^2$Department of Physics, National Taiwan University, Daan Taipei 10617, Taiwan\\
$^3$Laboratoire de Physique Th\'{e}orique, IRSAMC, Universit\'{e} de Toulouse, CNRS, UPS, France.\\
$^4$Institute of Physics, Academia Sinica, Nankang Taipei 11529, Taiwan\\
}

\date{\today}

\begin{abstract}
We study the incommensurate charge ordered states in the $\mathit{t-t^{\prime}-J}$ model using the Gutzwiller mean field theory on large systems. In particular, we explore the properties of  incommensurate charge modulated states referred to as nodal pair density waves (nPDW) in the literature. nPDW states intertwine site and bond charge order with modulated $d$-wave pair order, and are characterized by a nonzero amplitude of uniform  pairing; they also manifest a dominant intra-unit cell $d$-density wave form factor. To compare with a recent scanning tunneling microscopy (STM) study [Hamidian \textit{et al.}, Nat. Phys. 3519 (2015)] of the cuprate superconductor BSCCO-2212, we compute the continuum local density of states (LDOS) at a typical STM tip height using the Wannier function based approach. By Fourier transforming Cu and O sub-lattice LDOS we also obtain  bias-dependent intra-unit cell form factors and spatial phase difference. We find that in the nPDW state the behavior of form factors and spatial phase difference as a function of energy agrees remarkably well with the experiment.This is in contrast to  commensurate charge modulated states, which we show do not agree with experiment. We propose that the nPDW  states are good candidates for the charge density wave phase observed in the superconducting state of underdoped cuprates.
\end{abstract}


\maketitle

\section{Introduction}
Recent interest in cuprates has been spurred by the observation of charge order, in the underdoped regime of its phase diagram, through a variety of experimental tools like STM\cite{Koshaka07}, NMR \cite{Wu}, x-ray diffraction \cite{Chang}, and resonant x-ray scattering \cite{Damascelli}. Charge order appears to be a generic feature  of the phase diagram being observed in materials across the cuprate family such as YBCO \cite{Ghiringheli, Achkar1, Chang}, BSCCO \cite{SilvaNeto}, LaBSCCO \cite{Comin}, NaCCOC \cite{Koshaka07} and HBCO \cite{Taillefer}. In contrast to the stripe phase observed, e.g.  in LBCO \cite{Tranquada}, charge order in other cuprates does not generally accompany any magnetic order. These are short range, uni-directional and incommesurate charge modulations with wavevectors $[0, Q]$ or $[Q, 0]$ where $Q$ is around 0.3 in  reciprocal lattice units. 

Understanding the momentum - energy dependence of  local intra-unit cell density  form factors as a function of doping can give clues to the origin and interplay of charge order with other symmetry broken phases of cuprates. By directly imaging conductance at Cu and O sublattices, Hamidian et. al. have shown that at lower energies intra-unit cell modulations at O sites are in-phase, giving rise to a dominant $s'$-form factor; however, at higher energies,  where the signal reaches a peak intensity a $d$-form factor dominates. The authors of Ref. \onlinecite{Koshaka08}  associated the characteristic energy scale of the $d$-form factor modulation to the pseudogap scale. Furthermore, in the higher energy range where the $d$-form factor dominates, the states below and above the Fermi energy show a robust phase difference of $\pi$. 
In a recent development, the same group has also discovered the modulation of the Cooper-pair density in the charge order phase of BSCCO by using Josephson tunneling microscopy \cite{Hamidian2}. The energy dependence of the form factors, along with the observations of pair density waves (PDW), put constraints on  theories of charge order in cuprates.

Theoretical efforts to understand the physics of the charge order phase have  mostly focused on the hot-spot approximation or weak-coupling treatment of the spin-fermion model \cite{Sachdev1, Pepin, Chubukov1}. Here, charge order appears as a dFF bond order competing with d-wave superconductivity, and having wave vectors along diagonal \cite{Sachdev1} or horizontal or vertical axes of the Brillouin zone \cite{Chubukov1}. 
The PDW state has also been shown to emerge from this model \cite{Chubukov3} as a possible ground state.  However, in a strong coupling Eliashberg type treatment of the model, bond order was found to be suppressed for experimentally relevant parameters \cite{Mishra}.

From a more localized perspective, charge ordered states have been investigated in $\mathit{t-J}$ type models using simple mean field approximation, renormalized  mean-field theory (Gutzwiller approximation)\cite{Yang, WeiLin}, variational Monte Carlo \cite{Himeda,Raczkowski,Chou} and infinite projected pair entangled states methods \cite{Carboz}. In particular, renormalized mean field theory predicts several unidirectional and bidirectional charge ordered states with ground state energies nearly degenerate with the uniform superconducting state \cite{Yang,Raczkowski,WeiLin}. Of these, anti-phase charge density wave (AP-CDW) and nPDW have dominant $d$-form factor and exist in the doping range where charge order has been experimentally observed \cite{WeiLin}.

The AP-CDW is a charge order with commensurate wave vector e.g. $[0, .25]$ or $[.25, 0]$, that has been studied extensively in Ref. 11 and 23. These states have an accompanying superconducting order parameter that forms domains with opposite signs (hence, the name anti-phase (AP) charge density wave). The nPDW is an incommensurate charge order with wave vector $[0, Q]$ or $[Q, 0]$, where $Q \approx 0.3$. In addition to the modulating component, the pair field has a uniform d-wave component giving rise to nodal structure in density of states at low energies similar to the experimental observation \cite{Koshaka07,Koshaka08}. Thus the nPDW intertwines uniform superconductivity, pair density wave and charge order. In this paper, we will show that this state correctly captures the energy dependence of form factors and spatial phase difference as seen in the STM experiments on the superconducting underdoped cuprates \cite{Hamidian}. 

STM and resonant x-ray scattering experiments directly probe the intra-unit cell lattice sites and provide information about form factors \cite{Damascelli, Yazdani}. However, in most  previous theoretical works, the form factor is obtained by finding the renormalization of nearest neighbor hoppings between unit cells along x- and y-axes \cite{Sachdev1, Sachdev2, Chubukov1}. A three band model  approach introducing planar oxygen states can be utilized to resolve this issue, but the treatment of the no-double occupancy criterion is more complicated \cite{Atkinson}. To avoid a three band calculation, and yet be able to compute LDOS maps and form factors as in STM, we have utilized a Wannier function based approach that has been successfully used to explain STM conductance maps around impurities in BSCCO and FeSe \cite{Choubey, Kreisel}.

In this paper, we first self-consistently solve the $\mathit{t-t^{\prime}-J}$ model under Gutzwiller approximation with suitable initial conditions to obtain nPDW state.
Next, we compute the lattice Green's functions and then change the basis to continuum space, with Wannier functions   as the matrix elements of transformation, to obtain the continuum Green's function. After obtaining the Cu and O sublattice continuum LDOS at typical STM tip heights, we then calculate the form factors and show that the results for nPDW state are in very good  agreement with the STM experiment \cite{Hamidian}. We find  at low energies that the $s^{\prime}$-form factor has largest contribution; however, at higher energies the $d$-form factor dominates. Moreover, we show that the $d$-form factor modulations 
are in phase at low energies and become out of phase at the energy scale at which $d$-form factor becomes dominant. Furthermore, we show that the modulation of $d$-wave pair order is crucial to obtain the bias dependence of the form factors similar to that observed in the experiment.

\section{Method}
The $t-t'-J$ Hamiltonian on a 2-D lattice is given by

\begin{equation}
\begin{aligned}
H=-\sum_{i,j,\sigma}P_{G}t_{ij}(c^\dagger_{i\sigma}c_{j\sigma}+H.C.)P_{G} + J\sum_{\langle i,j\rangle}\mathbf{S}_{i} \cdot \mathbf{S}_{j},
\end{aligned}
\end{equation}
where $c^\dagger_{i\sigma}$ creates an electron at lattice site $i = (i_{x}, i_{y})$  with spin $\sigma = \pm$ and $\mathbf{S}_{i}$ is the spin operator at this site. $P_G=\prod_i(1-n_{i\uparrow}n_{i\downarrow})$ is the Gutzwiller projection operator, where $n_{i\sigma}=c^\dagger_{i\sigma}c_{i\sigma}$ represents the spin dependent number operator for site $i$. The hopping matrix element $t_{ij}$ is equal to $t$ and $t'$ if $i$ and $j$ are nearest and next nearest neighbor sites, respectively. $t'=-0.3t$ and $J=-0.3t$ are used in this work. When necessary to compare with the experimental bias scale, we take $t = 400$ meV.

The no double occupancy constraint can be treated in Gutzwiller approximation scheme \cite{Zhang} where the projection operator is replaced by Gutzwiller counting factors leading to the following renormalized Hamiltonian.
\begin{equation}
\begin{aligned}
H=&-\sum_{i,j,\sigma}g^t_{ij}t_{ij}(c^\dagger_{i\sigma}c_{j\sigma}+H.C.)\\
&+\sum_{\langle i,j\rangle}J\Bigg [ g^{s,z}_{ij}S^{s,z}_i S^{s,z}_j+g^{s,xy}_{ij}\Bigg(\frac{S^{+}_i S^{-}_j+S^{-}_i S^{+}_j}{2}\Bigg)\Bigg] \\
\end{aligned}
\end{equation}

For non-magnetic states the simplified Gutzwiller factors are given as \cite{Yang}
\begin{equation}
\begin{aligned}
&g^t_{ij}=g^t_{i\sigma}g^t_{j\sigma}\\
&g^t_{i}=\sqrt{\frac{2\delta_i}{1+\delta_i}} \\
&g^{s,xy}_{ij}=g^{s,xy}_i g^{s,xy}_j\\
&g^{s,xy}_i=\frac{2}{1+\delta_i}\\
&g^{s,z}_{ij}=g^{s,xy}_{ij}=g^{s}_{ij}
\end{aligned}
\end{equation}

The Gutzwiller factors depend on the local values of the hole density $\delta_i$, pair field $\Delta_{ij\sigma}^v$, and bond field $\chi_{ij\sigma}^{v}$. As in Ref. \onlinecite{WeiLin}, the  superscript ${v}$ indicates that these quantities are related to, but not same as the physical order parameters. These mean-fields are given as 
\begin{equation}
\begin{aligned}
&\Delta_{ij\sigma}^{v}=\sigma \langle\Psi_0 | c_{i\sigma}c_{j\bar{\sigma}} | \Psi_0 \rangle\\
&\chi_{ij\sigma}^{v}=\langle\Psi_0 | c^\dagger_{i\sigma}c_{j\sigma} |\Psi_0 \rangle\\
&\delta_i=1-\langle\Psi_0 | n_i |\Psi_0 \rangle
\end{aligned}
\end{equation}

where, $\bar{\sigma} = -\sigma$. The $d$-wave superconducting gap order parameter and the bond order along $x$ ($y$) direction are given as

\begin{equation}
\begin{aligned}
\Delta_{i} = \frac{1}{8}\sum_{\sigma} (& g^t_{i,i + \hat{x},\sigma}\Delta_{i,i + \hat{x}, \sigma}^{v} + g^t_{i,i - \hat{x},\sigma}\Delta_{i,i - \hat{x}, \sigma}^{v} \\
&-g^t_{i,i + \hat{y},\sigma}\Delta_{i,i + \hat{y}, \sigma}^v - g^t_{i,i - \hat{y},\sigma}\Delta_{i,i - \hat{y}, \sigma}^{v}) 
\end{aligned}
\end{equation}

\begin{equation}
\begin{aligned}
\chi_{i, i + \hat{x} (\hat{y})} = \frac{1}{2}\sum_{\sigma} g^t_{i,i + \hat{x}(\hat{y}) ,\sigma}\chi_{i,i + \hat{x} (\hat{y}), \sigma}^{v} + H.C. 
\end{aligned}
\end{equation}

Variational minimization of the ground state energy $E_{g} = \langle\Psi_0 | H |\Psi_0 \rangle$, with respect to the unprojected wavefunction $|\Psi_0 \rangle$ under the constraints of normalization and fixed total occupancy leads to the following mean-field Hamiltonian,

\begin{equation}
\begin{aligned}
H_{MF}=&\sum_{i,j,\sigma}\epsilon_{ij\sigma}c^\dagger_{i\sigma}c_{j\sigma}+H.C.\\
&+\sum_{\langle i,j\rangle}\sigma D^{\ast}_{ij}c_{i\sigma}c_{j\sigma}+H.C. - \sum_{i \sigma} \mu_{i \sigma}n_{i \sigma}, \\
\end{aligned}
\end{equation}
where
\begin{equation}
\begin{aligned}
&\epsilon_{ij\sigma}= -\frac{3}{4}Jg^{s}_{ij}\chi^{v \ast}_{ij\sigma}\delta_{\langle ij\rangle} - g^t_{(ij)\sigma}t_{(ij)} + \frac{\partial W}{\partial g^{s}_{ij}}\frac{\partial g^{s}_{ij}}{\partial \chi_{(ij)\sigma}^{v}}\\
&D^{\ast}_{\langle ij\rangle} = -\frac{3}{4}Jg^{s}_{ij}\Delta^{v\ast}_{\langle ij\rangle\sigma} + \frac{\partial W}{\partial g^{s}_{ij}}\frac{\partial g^{s}_{ij}}{\partial \Delta_{\langle ij\rangle\sigma}^{v}}\\
&\mu_{i \sigma} = \mu - 2\sum_{j}\left[\frac{\partial W}{\partial g^{s}_{ij}}\frac{\partial g^{s}_{ij}}{\partial n_{i\sigma}} + \frac{\partial W}{\partial g^{t}_{ij\sigma}}\frac{\partial g^{t}_{ij\sigma}}{\partial n_{i\sigma}}\right]\\
& W =  \langle\Psi_0 | H |\Psi_0 \rangle - \lambda(\langle\Psi_0|\Psi_0 \rangle - 1) - \mu(\sum_{i} n_{i} - N_{e}) \nonumber
\end{aligned}
\end{equation}
Here, $\delta_{\langle ij\rangle} = 1$ if $i$ and $j$ are nearest neighbors and $0$ otherwise. $\lambda$ and $\mu$ are Lagarange multipliers and $N_{e}$ is the total electron filling.

Since the charge order observed in most of the cuprates is unidirectional \cite{Damascelli}, we will focus only on realization of such states in the extended $\mathit{t-J}$ model. We assume that charge modulation is in the x-direction and exploit translational invariance in y-direction by switching to the ($i_{x}, k$) basis defined by following transformation.

\begin{equation}
c^{\dagger}_{i\sigma} = \frac{1}{\sqrt{N}}\sum_{k}c^{\dagger}_{i_{x}
}(k)e^{-ikR_{i_{y}}},
\end{equation}

where $N$ is the lattice dimension in y-direction, $R_{i_{y}}$ is the y-component of the lattice vector corresponding to site $i$, and $c_{i_{x}k}$ creates an electron with transverse momentum $k$, at the site $i_{x}$ in the 1D lattice. In this "collapsed 1D" representation, the mean-field Hamiltonian becomes

\begin{equation}
\begin{aligned}
H_{MF}=&\sum_{i_{x},j_{x}, k, \sigma}\epsilon_{i_{x}j_{x}\sigma}(k)c^\dagger_{i_{x}\sigma}(k)c_{j_{x}\sigma}(k)+H.C.\\
&+\sum_{i_{x},j_{x}, k}\sigma D^{\ast}_{i_{x}j_{x} \sigma}(k)c_{i_{x}\sigma}(k)c_{j_{x}\bar{\sigma}}(-k)+H.C. \\
&- \sum_{i_{x}, k, \sigma} \mu_{i_{x} \sigma}n_{i_{x} \sigma}(k), \\
\end{aligned}
\end{equation}
where
\begin{equation}
\epsilon_{i_{x}j_{x}\sigma}(k) = \sum_{i_{y}}\epsilon_{i_{x}i_{y}j_{x}0}e^{-ikR_{i_{y}}}. \nonumber
\end{equation}
A similar expression holds for $D_{i_{x}j_{x} \sigma}(k)$ and $\mu_{i_{x}} = \mu_{(i_{x}, 0)}$. The above Hamiltonian can be diagonalized by a spin-generalized Bogoliubov-de Gennes (BdG) transformation, which leads to the following BdG equations,
\begin{equation}
\begin{aligned}
\sum_{j} \left[\begin{array}{cc}
\xi_{ij\uparrow}(k) & -D_{i j\uparrow}(-k)\\ -D_{j i \uparrow}^{\ast}(k) & -\xi^{\ast}_{ij\downarrow}(-k)
\end{array} \right] \left[\begin{array}{c} u_{j\uparrow}^{n}(k)  \\ v_{j\downarrow}^{n}(k) \end{array} \right] = E_{n \uparrow}(k) \left[\begin{array}{c} u_{i \uparrow}^{n}(k)  \\ v_{i\downarrow}^{n}(k) \end{array} \right].
\end{aligned}
\label{Eq:BdG}
\end{equation}
Here, $\xi_{ij\sigma}(k) = \epsilon_{ij\sigma}(k) - \mu_{i \sigma}\delta_{ij}$. For simplicity of notation, we have replaced $i_{x}$, $j_{x}$ with $i$ and $j$ respectively. The mean fields are given following equations,
\begin{equation}
\begin{aligned}
&n_{i_{x} \uparrow} = \frac{1}{N}\sum_{nk}\lvert u^{n}_{i_{x} \uparrow}(k)\rvert^2f\left(E_{n\uparrow}(k)\right)\\
&n_{i_{x} \downarrow} = \frac{1}{N}\sum_{nk}\lvert v^{n}_{i_{x} \downarrow}(k)\rvert^2\left[ 1 - f\left(E_{n\uparrow}(-k)\right)\right]\\
&\Delta^{v}_{i_{x}j_{x}\uparrow}(k) =  \sum_{n}u^{n}_{i_{x} \uparrow}(k)v^{n\ast}_{j_{x} \downarrow}(k)\left[ 1 - f\left(E_{n\uparrow}(k)\right)\right]\\
&\chi^{v}_{i_{x}j_{x}\uparrow}(k) =  \sum_{n}u^{n \ast}_{i_{x} \uparrow}(k)u^{n}_{j_{x} \uparrow}(k)f\left(E_{n\uparrow}(k)\right)\\
&\chi^{v}_{i_{x}j_{x}\downarrow}(k) =  \sum_{n}v^{n}_{i_{x} \downarrow}(-k)v^{n\ast}_{j_{x} \downarrow}(-k)\left[ 1 - f\left(E_{n\uparrow}(-k)\right)\right],\\
\end{aligned}
\label{Eq:meanfields}
\end{equation}
where $f$ represents the Fermi function. Eq. (\ref{Eq:BdG})  and (\ref{Eq:meanfields}) are solved self-consistently until the desired accuracy is obtained. With converged values of the eigenfunctions, the Green's function matrix can be calculated using 
\begin{equation}
\begin{aligned}
&G_{ij\sigma}(\omega) = \frac{1}{N}\sum_{k}g^{t}_{i}g^{t}_{j}G_{i_{x}j_{x}\sigma}(k,\omega)e^{ik\left(R_{i_{y}} - R_{j_{y}}\right)}\\
&G_{i_{x}j_{x}\sigma}(k,\omega) = \sum_{n>0}\left[\frac{u^{n}_{i_{x} \sigma}(k)u^{n\ast}_{j_{x}\sigma}(k)}{\omega - E_{n\sigma} + i0^{+}} + \frac{v^{n\ast}_{i_{x} \sigma}(k)v^{n}_{j_{x}\sigma}(k)}{\omega + E_{n\bar{\sigma}}(k) + i0^{+}}\right].
\end{aligned}
\label{Eq:latticeG}
\end{equation}
 Throughout this paper we have used the artificial broadening $0^{+} = 0.01t$. To compute the LDOS at the STM tip position, 
we change the basis and obtain the continuum Green's function using \cite{Choubey}.
\begin{equation}
\begin{aligned}
G_{\sigma}(\mathbf{r}, \omega) = \sum_{ij}G_{ij\sigma}(\omega)W_{i}(\mathbf{r})W^{\ast}_{j}(\mathbf{r}),
\end{aligned}
\label{Eq:continuumG}
\end{equation}
where $W_{i}(\mathbf{r})$ is the Wannier function at site $i$, and $\mathbf{r}$ is a three dimensional continuum real space vector. The Wannier function employed in this paper was generated using the Wannier90 package \cite{Mostofi}, and is similar in  form to that used in Ref. \onlinecite{Kreisel}.  Note that the local Green's function contains nonlocal contributions from all lattice sites.  The continuum local density of states is now easily obtained as
\begin{equation}
\begin{aligned}
\rho_{\sigma}(\mathbf{r}, \omega) = -\frac{1}{\pi}Im[G_{\sigma}(\mathbf{r}, \omega)]
\end{aligned}
\label{Eq:continuumLDOS}
\end{equation}

In most of the previous theoretical works \cite{Sachdev1,Sachdev2,Chubukov1}, intra-unit cell form factors were calculated using the Fourier transform of the  nearest neighbor bond order $\chi_{i,i + \hat{x}(\hat{y})}$, which can be regarded as the measure of charge density at the oxygen atoms on $x(y)$ bonds at lattice site $i$. We can express $s$-, $s^{\prime}$-, and $d$-form factors as follows.

\begin{equation}
\begin{aligned}
&{D}^{\chi}(\textbf{q})= FT(\tilde{\chi}_{i,i + \hat{x}} - \tilde{\chi}_{i,i + \hat{y}})/2 \\
&{S'}^{\chi}(\textbf{q})= FT(\tilde{\chi}_{i,i + \hat{x}} + \tilde{\chi}_{i,i + \hat{y}})/2 \\
&{S}^{\chi}(\textbf{q})= FT(1 - \tilde{\delta}_{i}), \\
%
\end{aligned}
\label{Eq:formfactors_BondOrder}
\end{equation}
where $FT$ refers to the Fourier transform and $\tilde{}$ denotes that the spatial average of the corresponding quantity has been subtracted to emphasize modulating components. Obviously, this quantity does not have any energy dependence. However, STM experiments utilized phase resolved sublattice LDOS information \cite{Fujita} to extract the form factors and found a significant bias dependence\cite{Hamidian}. Using the continuum LDOS information, we can follow a similar approach. First, we obtain LDOS Z-maps, defined below, on a plane located at a typical STM tip height ($\approx 5 $ \AA) above the BiO plane.

\begin{equation}
\begin{aligned}
\rho^{Z}(\mathbf{r}, \omega>0) = \frac{\sum_{\sigma}\rho_{\sigma}(\mathbf{r}, \omega)}{\sum_{\sigma}\rho_{\sigma}(\mathbf{r}, -\omega)}
\end{aligned}
\label{Eq:continuumZ}
\end{equation}

Next, we take non-overlapping square regions around each atom in the Z-map, with the size of the region identical to that used in the experiment \cite{Hamidian, Davis}, and subsequently assign it to the sublattice Z-maps $Cu^{Z}(\mathbf{r},\omega)$, $O^{Z}_{x}(\mathbf{r},\omega)$ and $O^{Z}_{y}(\mathbf{r},\omega)$. We note that form factor results are not very sensitive to the size of the square region, however. Here subscripts x and y designate two nonequivalent oxygen atoms in the unit cell in horizontal and vertical directions, respectively.  Taking the proper linear combination of the Fourier transform of the sublattice LDOS yields $s$-, $s^{\prime}$-, and $d$-form factors as follows. 

\begin{equation}
\begin{aligned}
&{D}^Z(\textbf{q}, \omega )=(\tilde{{O}}^Z_x(\textbf{q}, \omega )- \tilde{{O}}^Z_y(\textbf{q}, \omega ))/2 \\
&{S'}^Z(\textbf{q}, \omega )=(\tilde{{O}}^Z_x(\textbf{q}, \omega )+\tilde{{O}}^Z_y(\textbf{q}, \omega ))/2 \\
&{S}^Z(\textbf{q}, \omega )=\tilde{{Cu}}^Z(\textbf{q}, \omega )
\end{aligned}
\label{Eq:formfactors}
\end{equation}

Another important quantity of interest is the average spatial phase difference ($\Delta\phi$) between the positive and negative bias energies for the $d$-form factor modulations. To compute $\Delta\phi$ in accordance with the experimental procedure \cite{Hamidian}, we filter out the characteristic wave vector corresponding to d-form factor modulation (${\mathbf{Q}}_{d}$) from the continuum LDOS maps at positive and negative energies using a Gaussian filter. Then we take the inverse Fourier transform to obtain the complex spatial map $D(\mathbf{r}, \omega)$ and determine its phase $\phi(\mathbf{r}, \omega)$. By taking the average of the spatial phase difference at $\pm \omega$, we find $\Delta\phi$.

\begin{equation}
\begin{aligned}
&{D^{g}}(\textbf{q},\omega) =  (\tilde{{O}}^g_x(\textbf{q}, \omega )- \tilde{{O}}^g_y(\textbf{q}, \omega ))/2 \\
&D(\mathbf{r},\omega)=\frac{2}{(2\pi)^2}\int d\mathbf{q} e^{i\textbf{q}\textbf{r}}{D^{g}}(\textbf{q},\omega) e^{-\frac{(\textbf{q}-{\mathbf{Q}}_{d})^2}{2\Lambda^2}} \\
&\phi(\mathbf{r}, \omega) = \arctan\left({\rm Im}[D(\mathbf{r},\omega)]/{\rm Re}[D(\mathbf{r},\omega)]\right)\\
& \Delta\phi = \langle \phi(\mathbf{r}, \omega) - \phi(\mathbf{r}, -\omega)\rangle,
\end{aligned}
\label{Eq:phase}
\end{equation}

where $\tilde{O}^g_x({\bf q},\omega)$ and $\tilde{O}^g_y({\bf q},\omega)$ are the Fourier transforms of the sublattice LDOS maps for oxygen x and oxygen y. Width of the Guassian filter was taken to be $\Lambda = 1/2N$.

 The full computational procedure is summarized as follows. First, we start with trial values for the hole density on each site, bond field on nearest and next nearest neighbor sites and pair field on nearest neighbor sites in a 2D lattice comprising N$\times$N lattice sites. The hole density and bond field are taken to be uniform, whereas the trial pair field is assumed to have a sinusoidal modulation with a given amplitude $\Delta_{0}$ and wave number $Q_{0}$. With this initial guess, the  BdG equations (Eqs. \ref{Eq:BdG} and \ref{Eq:meanfields}) are solved self-consistently until a converged solution is obtained. Then we find the Green's function matrix using Eq. (\ref{Eq:latticeG}) in a supercell set-up to gain higher energy resolution. Finally, we obtain continuum LDOS at height $\approx 5$\AA~ above BiO plane using Eqs. (\ref{Eq:continuumG})-(\ref{Eq:continuumZ}), and compute form factors and spatial phase difference using Eqs. (\ref{Eq:formfactors}) and (\ref{Eq:phase}),  respectively.


\section{Results}

\begin{figure}
\includegraphics[width=.7\columnwidth]{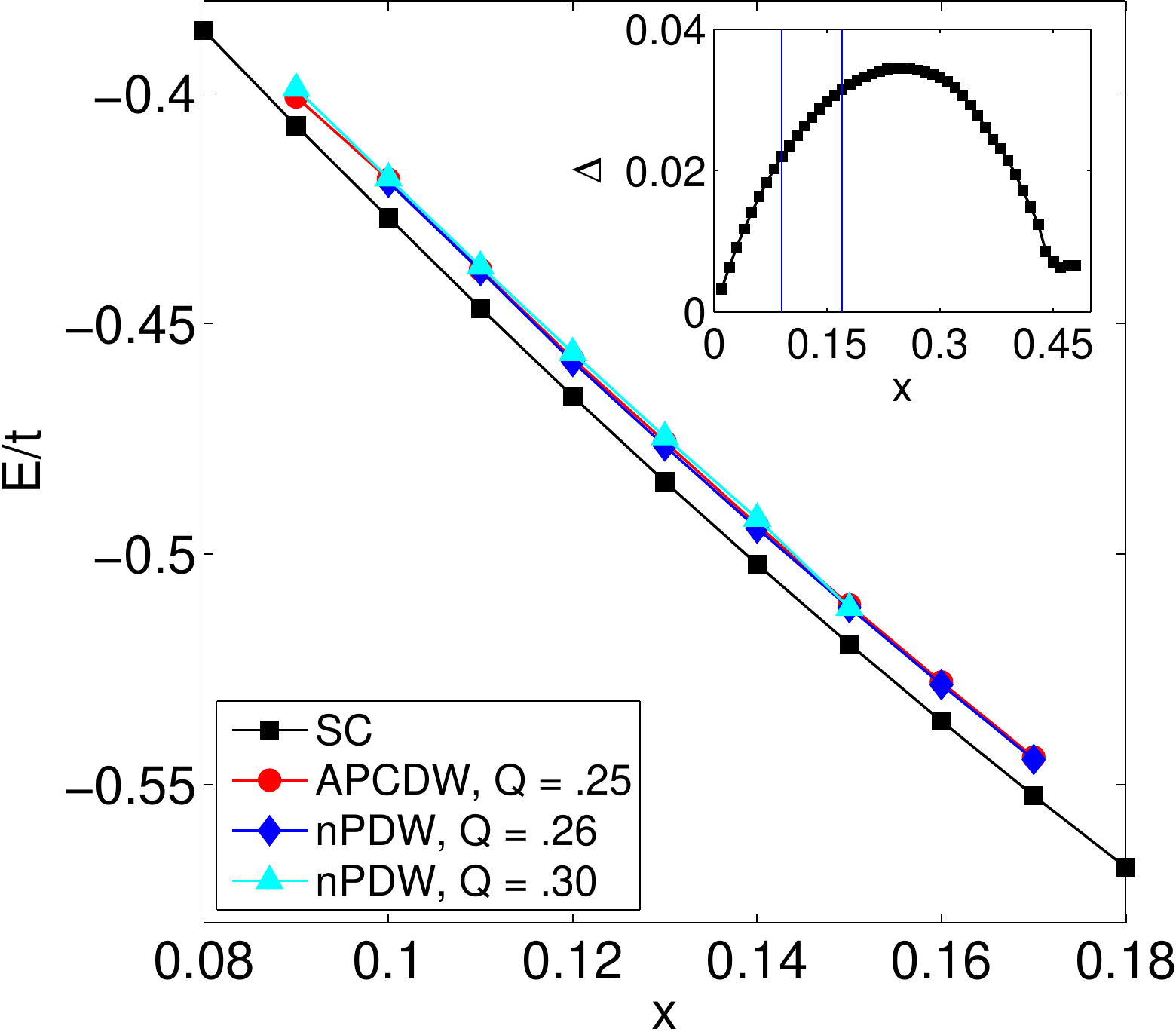}
\caption{(Color online) Energy per site (E/t) at various hole dopings (x) for homogeneous superconducting state, APCDW, and nPDW states. Inset: Variation of superconducting order parameter in the homogeneous state as a function of hole doping. Vertical lines mark the doping range in which APCDW and nPDW states are realized.}
\label{fig:Eg_vs_doping}
\end{figure}

Fig. \ref{fig:Eg_vs_doping} shows the plot of the energy per site ($E/t$) for the uniform superconducting and the charge ordered states as a function of hole doping. In the inset we plot the gap order parameter in the uniform superconducting state in doping range 0.01-0.48 where it is realized in $\mathit{t-t^{\prime}-J}$ model for the parameters considered.  It is well-known that the Gutzwiller approximation to the homogeneous $t-t'-J$ phase diagram is similar to cuprates, but with a renormalized doping scale. APCDW and nPDW states were found in the doping range of 0.09-0.17 which is below the optimal hole doping of 0.27. We note that in BSCCO charge order is found empirically vanishes at a critical doping $x_{c} \approx 0.19$ which is slightly above the optimal hole doping \cite{Fujita2}. 

It is clear from Fig. \ref{fig:Eg_vs_doping} that the uniform superconducting state has lower energy per unit site than the charge ordered states. A similar conclusion has been reached in previous studies using variety of numerical techniques \cite{WeiLin,Yang,Carboz,Raczkowski}. We will return to this aspect in section \ref{Discussion}, where we discuss various scenarios in which the energy of charge ordered states can be lowered relative to the uniform superconducting state. Another striking feature is that the APCDW and, nPDW states with different wave vectors are very close in energy at each and every hole doping. 
 This is consistent with the previous study \cite{WeiLin} where authors find a large number of nearly degenerate inhomogeneous solutions, although they considered a smaller $16\times16$ lattice and $t^{\prime}=0$.

In following subsections we describe each charge ordered state in detail.

\subsection{APCDW}
\begin{figure}
\includegraphics[width=1\columnwidth]{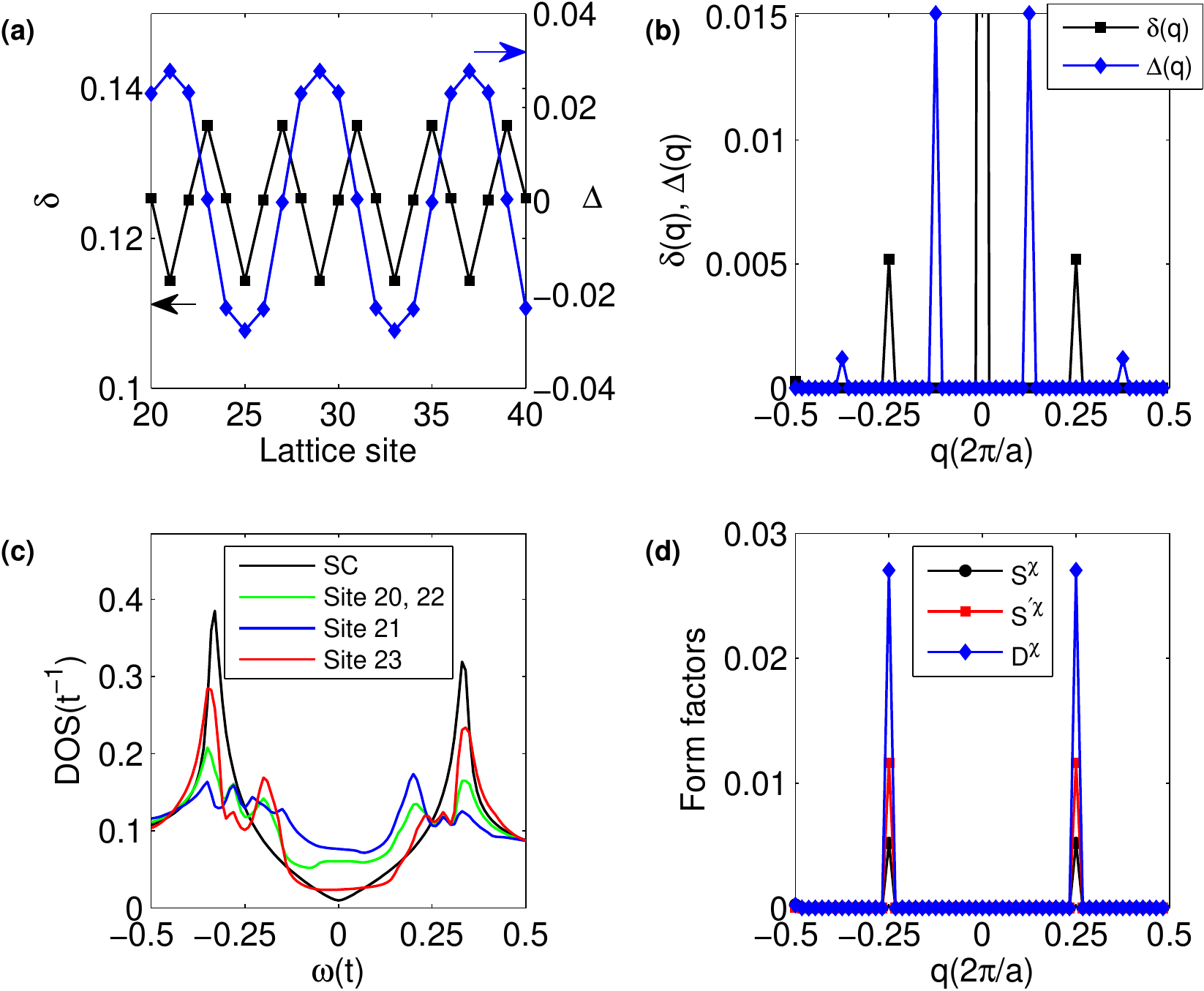}
\caption{ (Color online) Characteristic features of APCDW state. (a) Variation of hole density ($\delta$) and gap order parameter ($\Delta$) with lattice sites in the central region of $56\times56$ system. y-axis in left (right) corresponds to $\delta$ ($\Delta$). (b) Fourier transform of the hole density ($\delta(q)$) and gap order parameter ($\Delta(q)$). The $q = 0$ component of hole density modulation, not shown in the plot, is 0.125. (c) Density of states in the homogeneous superconducting state and, APCDW state over a period of lattice sites. (d) Intra-unit cell form factors in APCDW state computed using Eq. \ref{Eq:formfactors_BondOrder}.}
\label{fig:APCDW_details}
\end{figure}

We first present the results for commensurate APCDW state to set the stage for the discussion of the more complicated incommensurate nPDW phase.
The APCDW state has been investigated in Refs. \onlinecite{WeiLin, Yang} (Note that  Yang et. al. \cite{Yang} referred to it as the $\pi$DW state). To get an APCDW state, with a periodicity of four lattice constants, we work with a system size which is multiple of 8 and, initialize the BdG equations (\ref{Eq:BdG}) and (\ref{Eq:meanfields}) with a pair field modulating at wave number $Q_{0} = 1/8$. Results are shown in Fig. \ref{fig:APCDW_details} for a $56\times56$ system at the hole doping $x = 0.125$.

Fig. \ref{fig:APCDW_details}(a) shows the variation of hole density ($\delta$) and superconducting gap order parameter ($\Delta$) with lattice sites in the central region of the $56\times56$ system. The wavelength of the gap modulation is 8 which is twice of the wavelength of the charge modulation. Moreover, the gap changes sign after each period of the charge modulation, and the hole density is found to be maximum at the domain wall sites where gap order parameter vanishes. As is evident from Fig. \ref{fig:APCDW_details}(b), these real space observations are reflected in the Fourier domain where we find dominant charge modulation wave vector ($Q$) and gap modulation wave vector ($Q_{\Delta}$) to be 0.25 and 0.125 respectively. Also, the uniform component of the gap is zero. Thus, the APCDW phase intertwines a unidirectional pair density wave and charge density wave with wave vectors differing by a factor of two. Our result for $t^{\prime} = -0.3$ is similar to that of $t^{\prime} = 0$ in Ref. 11.
 
The local density of states (LDOS) on lattice sites over a period of APCDW state is plotted in Fig. \ref{fig:APCDW_details}(c) along with the LDOS in homogeneous state. The APCDW LDOS is finite at all lattice sites, and exhibits two sets of coherence peaks with higher energy peak almost coinciding with homogeneous state coherence peaks. The energy peaks are attributed to overlapping Andreev bound states (ABS) which appear at the domain wall sites where superconducting order parameter changes sign \cite{Yang}. The ABS form a one-dimensional band, and their overlap between neighboring domain walls broadens and shifts the ABS energy away from the chemical potential. 
Similar conclusions were drawn in connection to the Fulde-Ferrel-Larkin-Ovchinikov (FFLO) phase \cite{Vorontsov}.       

In Fig. \ref{fig:APCDW_details}(d) we show the density wave form factors computed from the bond order on nearest neighbor using Eq. \ref{Eq:formfactors_BondOrder}. Clearly, APCDW state exhibits dominant $d$-form factor at the wave vector $[0.25, 0]$.  However, the detailed bias 
dependence of these states do not match experimental results well.  We discuss these comparisons in the Supplementary Information.

\subsection{nPDW}
It is well established that the charge order observed in most of the cuprates is incommensurate and, generally, does not accompany a magnetic order \cite{Damascelli}. Thus, in order to address experiments in the context of $\mathit{t-t^{\prime}-J}$ model, we must look for  incommensurate order. However, a truly incommensurate charge order can not be realized in a finite lattice calculation. Instead, we can get a quasi-incommensurate state as a mixture of charge ordered states having different commensurate ordering wave vectors. To achieve this, we initialize the BdG equations (\ref{Eq:BdG}) and (\ref{Eq:meanfields}) with a pair field modulating at wave number which is not a multiple of $1/N$, i.e. $Q_{0} \neq \frac{m}{N}$ where, $m$ is an integer. Such assignment of wave vector ensures that the initial seed to BdG equations has more than one Fourier component, and thus a self-consistent solution may converge to a quasi-incommensurate state. The nPDW state, first reported in Ref. 11, was obtained using a slightly different initialization procedure. In the following we show the results obtained for a 60x60 system with initial wave number guess $Q_{0} = 0.154$. 
\subsubsection{Characterization of nPDW}

\begin{figure}
\includegraphics[width=1\columnwidth]{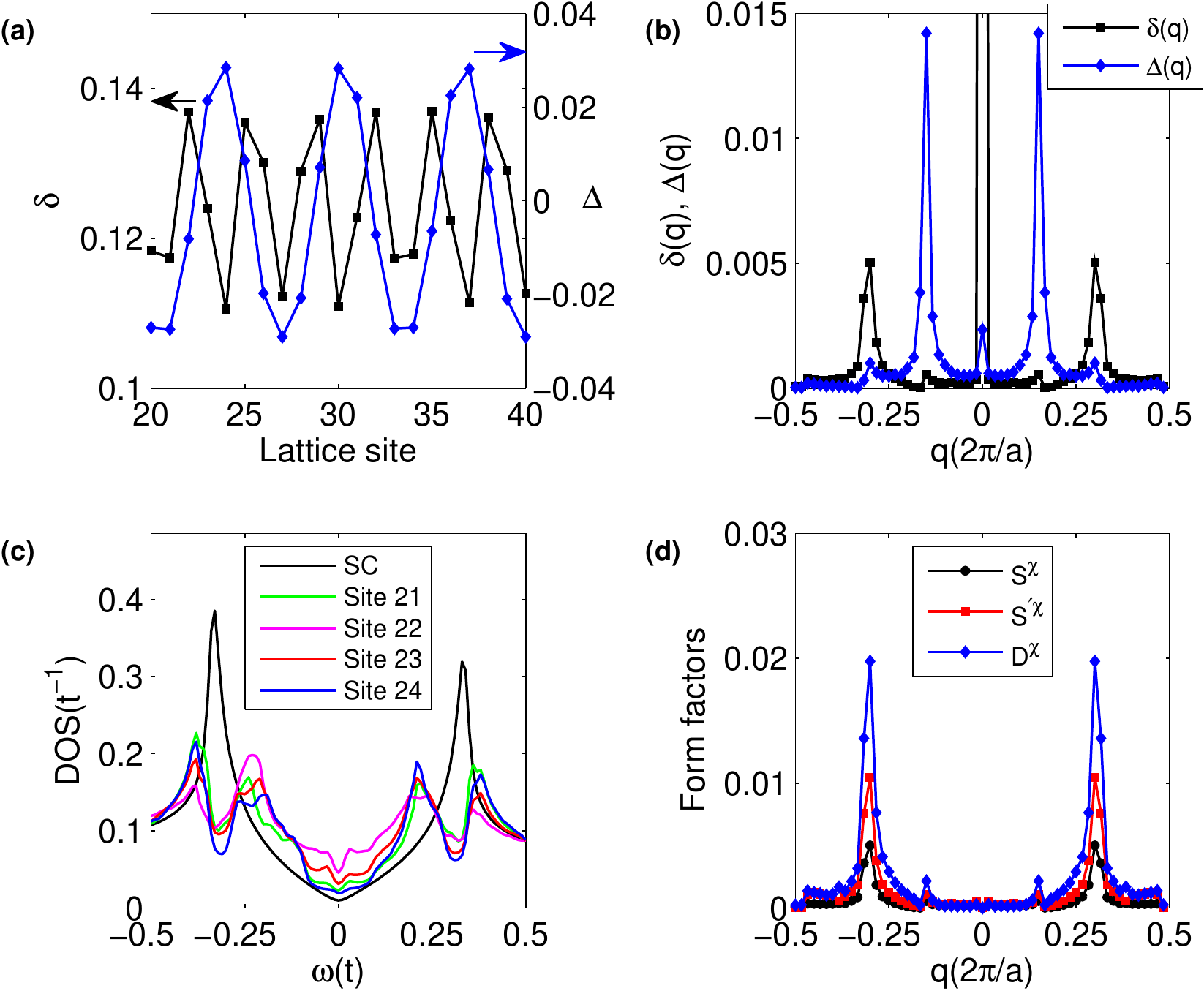}
\caption{ (Color online) Characteristic features of nPDW state. (a) Variation of hole density ($\delta$) and gap order parameter ($\Delta$) with lattice sites in the central region of $60\times60$ system. y-axis on left (right) corresponds to $\delta$ ($\Delta$). (b) Fourier transform of the hole density ($\delta(q)$) and gap order parameter ($\Delta(q)$). The $q = 0$ component of hole density modulation, not shown in the plot, is 0.125. (c) Density of states in the homogeneous superconducting state and nPDW state on four consecutive lattice sites. (d) Intra-unit cell form factors in nPDW state computed using Eq. \ref{Eq:formfactors_BondOrder}.}
\label{fig:nPDW_details}
\end{figure}

Fig. \ref{fig:nPDW_details} shows the characteristic features of the nPDW state in real and Fourier space. Fig.  \ref{fig:nPDW_details}(a) shows the variation of hole density and gap order parameter in the central region of 60x60 system at a hole doping of 0.125. 
The root mean square variation in hole density is found to be $\Delta \delta_{rms} = 0.01$ holes. Interestingly, an NMR experiment \cite{Wu} on the charge ordered phase of YBCO, with hole doping of 0.108, finds a charge density variation $\Delta \delta = 0.03 \pm 0.01$ holes which is of the same order as our findings. Like the case of the APCDW here too, the hole density is maximum at the domain wall site which, in this case, is defined as the lattice site where the order parameter changes sign. The quasi-incommensurate nature of the nPDW is reflected in Fig. \ref{fig:nPDW_details}(b) which shows that the hole density and the order parameter have many Fourier components. The dominant Fourier components in the order parameter ($Q_{\Delta}$)  and hole density ($Q$) are $0.15$ and $0.3$ respectively, satisfying $Q = 2Q_{\Delta}$ as in the APCDW case.

Another characteristic feature of the nPDW is the finite uniform component of the gap order parameter which results into non-vanishing d-wave global pairing as opposed to the case of APCDW. Thus, the nPDW state intertwines quasi-incommensurate charge density wave, pair density wave, and uniform d-wave superconductivity. Fig. \ref{fig:nPDW_details}(c) shows the density of states on a number of lattice sites in nPDW state. Like APCDW, there are two sets of coherence peaks. Higher energy peaks are slightly shifted away from the coherence peak in the uniform superconducting state. Lower energy peaks can be attributed to the hybridization of Andreev bound states arising due to domains of sign changing pair field. More importantly, the DOS has a nodal structure around Fermi energy at all lattice sites. This is a direct consequence of the non-vanishing global $d$-wave pairing. Fig. \ref{fig:nPDW_details}(d) shows that the form factors computed from bond order parameter have several Fourier components with $Q_{\chi} = 0.3$  the dominant one. Here too, the $d$-form factor has  the largest weight.  

\subsubsection{Continuum LDOS}
\begin{figure}
\includegraphics[width=.75\columnwidth]{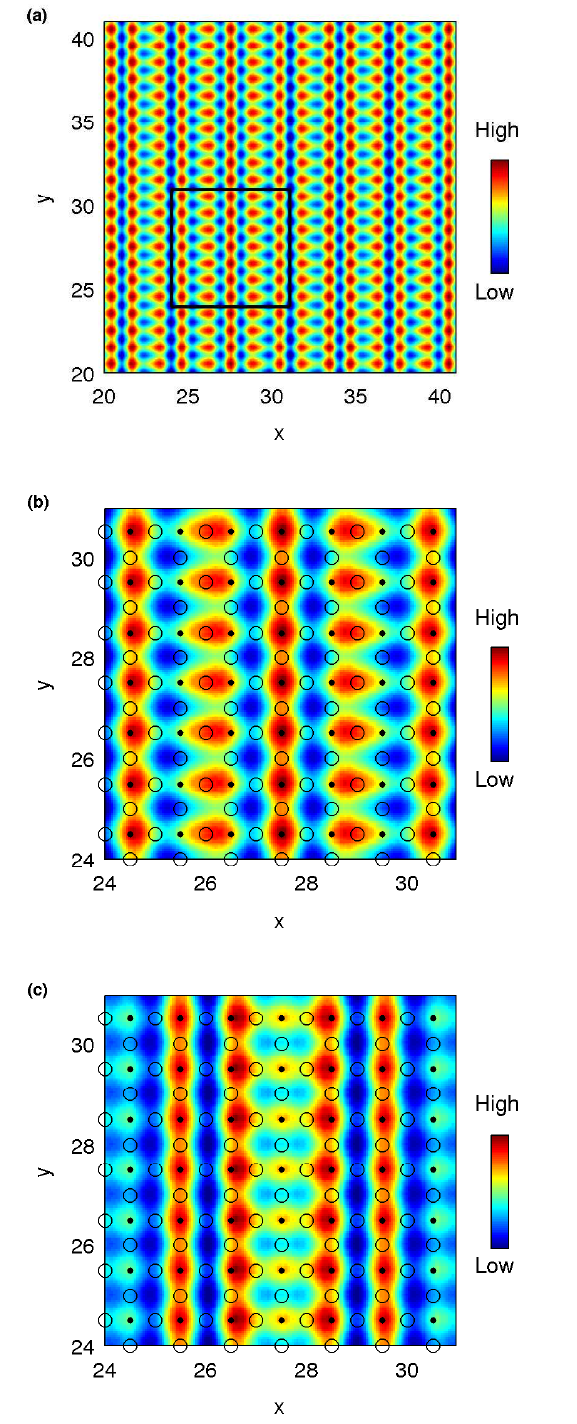}
\caption{ (Color online) Continuum LDOS map at $\omega = \pm 0.25t$ and $\approx$ 5 {\AA} above BiO plane. (a) LDOS map at $\omega = 0.25t$ in a range of 20x20 unit cells. (b) Zoomed-in view of the area marked by square in (a). Black dots and open circles represent position of Cu and O atoms, respectively, in the CuO plane underneath. (c) LDOS map at $\omega = - 0.25t$ in the same region as in (b).}
\label{fig:nPDW_ldosMap}
\end{figure}

For a more fruitful comparison with STM experiment, we turn to the continuum LDOS and quantities derived from it. With the first-principles Wannier function for BSCCO-2212 \cite{Kreisel} as an input, we compute the continuum LDOS using Eq. \ref{Eq:continuumG} and \ref{Eq:continuumLDOS}. The resulting LDOS map at energy $\omega = 0.25t$ and in a 20x20 unit cell area at a height $z \approx 5$ {\AA} above BiO plane, which is a typical height for STM tip, is shown in Fig. \ref{fig:nPDW_ldosMap}(a). Two types of modulating stripe structures can be observed. In Fig. \ref{fig:nPDW_ldosMap}(b) we plot a zoomed-in view of one of these structures in the area bounded by a square as shown in 4(a). Cu and O atoms located in the CuO plane underneath, are represented as dots and open circles, respectively. The LDOS shows modulations around all atoms which, in the Fourier domain, implies that this particular bias has a mixture of all intra-unit cell form factors.

More importantly, modulations at the two inequivalent O atoms in an unit cell ($\textrm{O}_{x}$ and $\textrm{O}_{y}$) are out of phase, i.e. when $\textrm{O}_{x}$ has large LDOS then $O_{y}$ has small LDOS. This leads to the conclusion that the $d$-form factor has larger weight than $s^{\prime}$-form factor. Indeed, a more quantitative analysis of form factors, discussed in following paragraphs, shows that the $d$-form factor has largest weight at this particular bias. This particular pattern is observed in an energy range of $0.21t-0.27t$. Remarkably, similar pattern has been observed in the STM experiments \cite{Fujita, Hamidian}. In Fig. \ref{fig:nPDW_ldosMap}(c) LDOS map is plotted at the negative bias $\omega = -0.25t$ in the same region as in (b). Comparing Fig. \ref{fig:nPDW_ldosMap}(b) and (c), it is found that the atoms with larger values of LDOS at positive bias have smaller values  at negative bias, which implies a spatial phase change of $\pi$ between positive and negative biases. As emphasized in Ref. [9], this is a characteristic feature of $d$-form factor density wave. A more quantitative analysis of the phase differences, calculated using Eg. \ref{Eq:phase}, is given in following paragraphs.

\subsubsection{Bias and doping dependence}

\begin{figure}
\includegraphics[width=1\columnwidth]{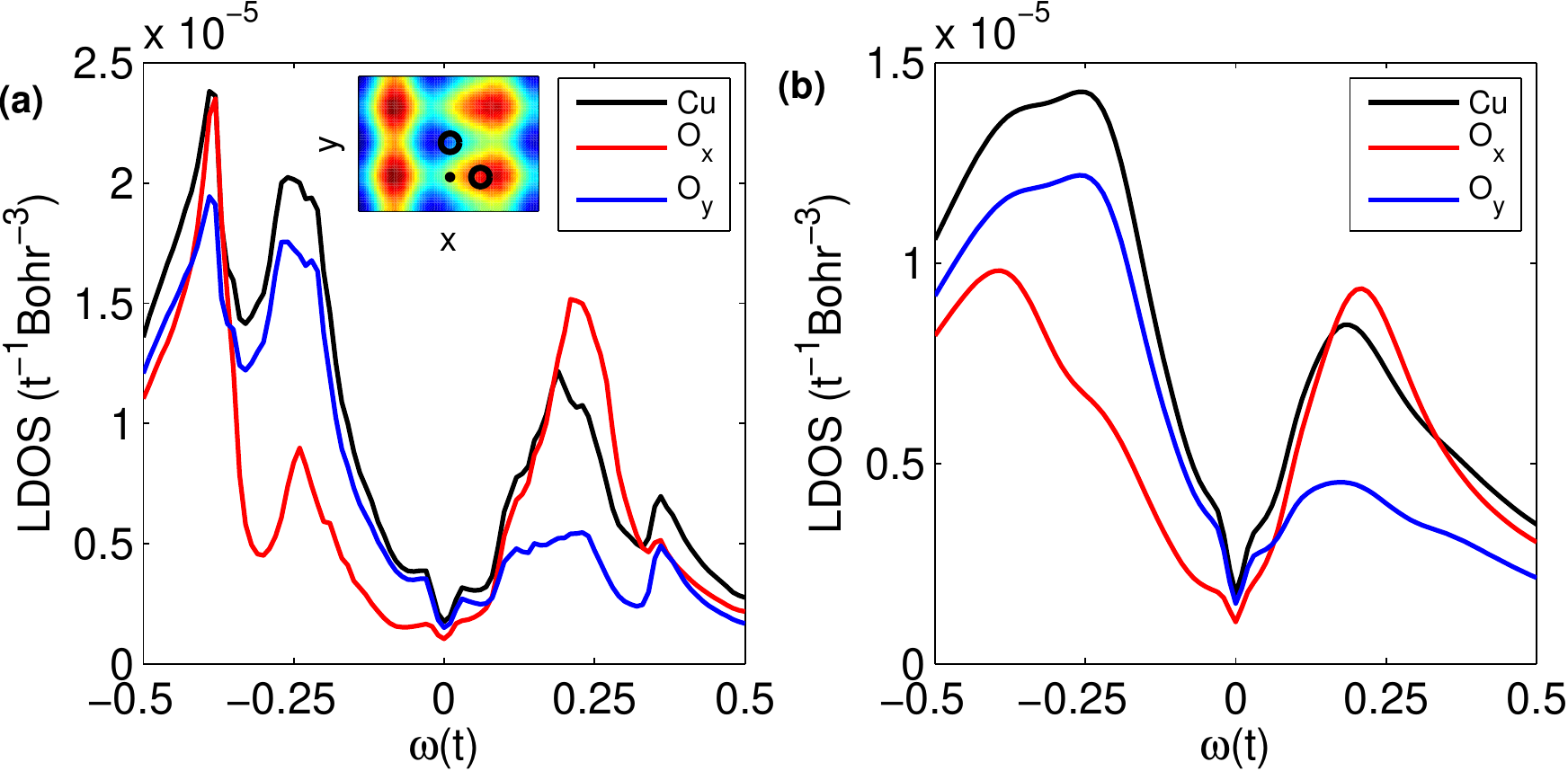}
\caption{ (Color online) Continuum LDOS spectrum 5 \AA above BiO plane registered above Cu, $\textrm{O}_{x}$ and $\textrm{O}_{y}$ sites in the unit cell [25, 25] at an height $\approx$ 5 {\AA} above BiO plane (a) without, and (b) with $\Gamma = \alpha \vert\omega\vert$ inelastic scattering ($\alpha$ = 0.25), as extracted in Ref.\onlinecite{Alldredge}. The location of the unit cell can be referred from Fig. \ref{Eq:continuumLDOS}(b) as shown in the inset. Dot and open circles represent Cu and O atoms respectively.}
\label{fig:npdw_ldos_spectrum}
\end{figure}

Fig. \ref{fig:npdw_ldos_spectrum}(a) shows the bias dependence of continuum LDOS at Cu, $\textrm{O}_{x}$, and $\textrm{O}_{y}$ atomic positions in unit cell [25, 25] of a $60\times 60$ system located at a height $\approx 5$ {\AA} above the surface BiO plane. The location of this particular unit cell in reference to others can be found in the lower left corner of Fig. \ref{fig:nPDW_ldosMap}(b), and the cell is shown explicitly in the inset of Fig. \ref{fig:npdw_ldos_spectrum}. Similar to the lattice LDOS, two sets of "coherence peaks" at $\approx \pm 0.21t$ and $0.37t$ can be observed. These peaks correspond to the modulated Andreev state created by the PDW and that associated with the charge density wave energy scale\cite{Yang}.

Also, a small gap-like feature exists around around the Fermi level due to the uniform component of the gap order parameter. The most striking feature is the difference between the LDOS at $\textrm{O}_{x}$ and $\textrm{O}_{y}$ atoms, which clearly demonstrates intra-unit cell $C_4$ symmetry breaking. The difference between the  two is maximum at $\omega \approx \pm 0.21t$, the scale corresponding to the hybridized Andreev bound states. As we will see, this is the bias at which $d$-form factor has largest magnitude. Another feature of the LDOS is the strong particle-hole asymmetry. Interestingly, this asymmetry is seen to be much more pronounced in the continuum LDOS than lattice LDOS (Fig. \ref{fig:nPDW_details}(c)). 

We expect that these intra-unit cell contrast of these various effects will be mitigated somewhat when nonzero tip size is accounted for\cite{AndreasLiFeAs1}.  In addition, we expect the higher-energy features to be broadened significantly by inelastic scattering\cite{Alldredge}. To see this effect explicitly, we incorporate linear inelastic scattering by replacing the constant artificial broadening term ($i0^{+}$) in Eq. \ref{Eq:latticeG} by an energy dependent artificial broadening $i0^{+} + i\Gamma(\omega)$ where  $\Gamma(\omega) = \alpha \vert\omega\vert$, as observed in Ref. \onlinecite{Alldredge}. Fig. \ref{fig:npdw_ldos_spectrum}(b) shows the resulting continuum LDOS spectrum for $\alpha = 0.25$
. We find that high energy peaks are indeed broadened and can not be resolved any more. This holds for all higher values of $\alpha$. The spectrum resemble those taken on O$_x$,  O$_y$, and Cu sites very closely\cite{Koshaka07}. We note that the value of spectral gap in BSCCO-2212 spectrum reported in Ref. 12 is in the range 80-90 meV for which the value of $\alpha$ is found to be in the range 0.25-0.33, justifying our choice \cite{Alldredge}.

 Using the continuum LDOS map, we can calculate energy dependent form factors as formulated in Eq. \ref{Eq:formfactors}. 
 To calculate the wave vector corresponding to $d$-form factor modulation ($\mathbf{Q}_d$), we compute the $d$-form factor ($D^{Z}(\mathbf{q}, \omega)$) as a function of energy and obtain the wave vector at which it peaks. We find that above a threshold bias,  this wave vector does not show any dispersion and remains constant at $\mathbf{Q}_d = [0.3, 0]$. This non-dispersing behavior is very similar to that seen in the experiment \cite{Hamidian}. 

\begin{figure}
\includegraphics[width=.6\columnwidth]{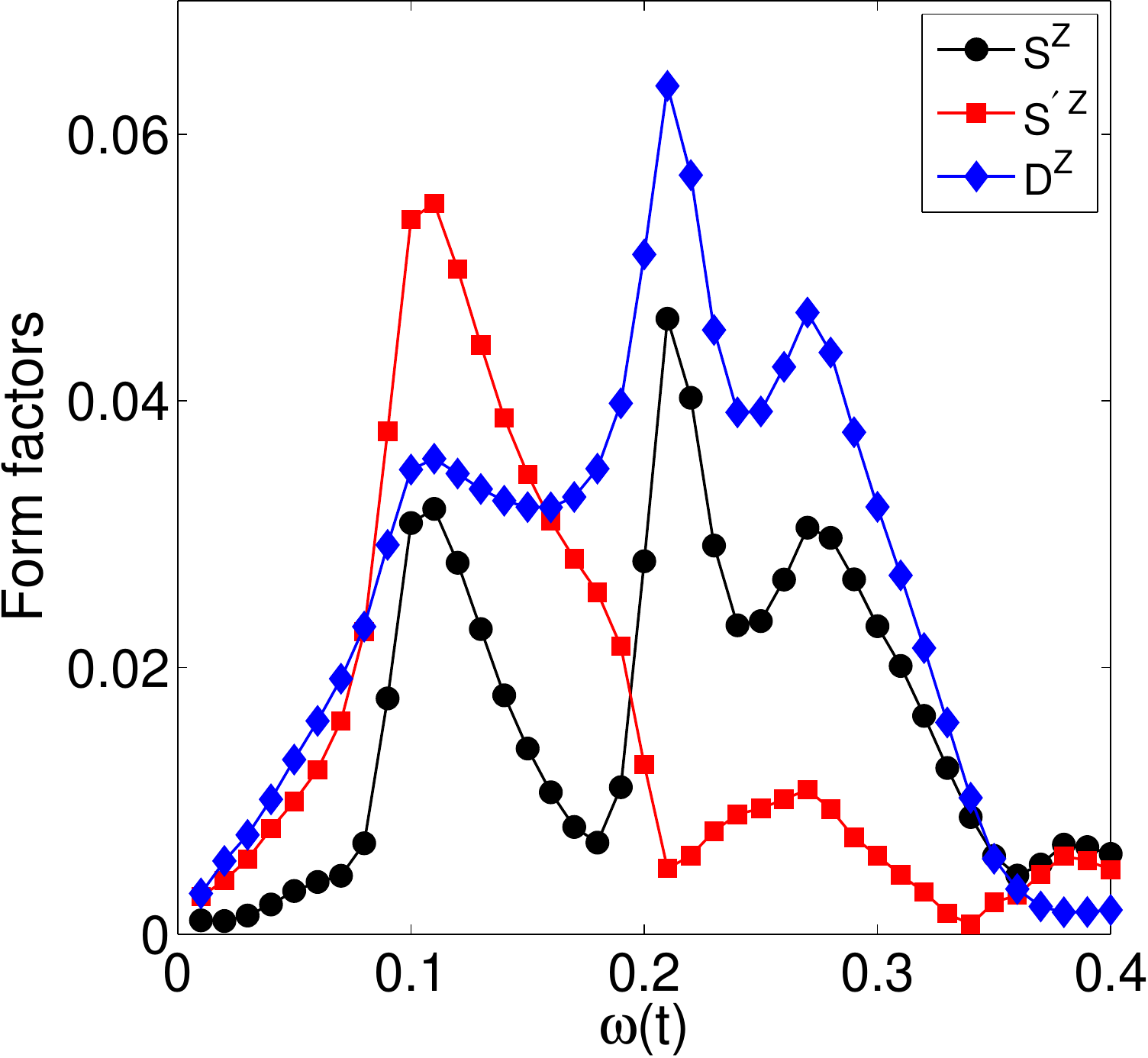}
\caption{ (Color online) Bias dependence of the intra-unit cell form factors at $x=0.125$ computed from atomic sublattice averages as described in the text. }
\label{fig:nPDW_formFactors}
\end{figure}

The energy dependence of the form factors at wave vector $\mathbf{Q}_d = [0.3, 0]$ is now shown in Fig. \ref{fig:nPDW_formFactors}. Similar to the experiment, we find an $s'$-form factor peak at a lower energy and a $d$-form factor peak at higher energy. Comparing the energy scales in the lattice LDOS (Fig. \ref{fig:nPDW_details}(c)) and continuum LDOS (Fig. \ref{fig:npdw_ldos_spectrum}), we find that the energy at which $d$-form factor peaks ($\Omega_{d}$), corresponds to the Andreev bound state peak. By studying the bias dependence of form factors in systems with varying $t'$, doping level and modulation wave vectors, we find that the $d$-form factor always displays a peak and the particular bias at which it occurs corresponds to the Andreev bound state peak in the lattice LDOS. However, the relative weight of the $s'$- and $d$-form factor depends on the details of band structure and doping. For example if we choose $t' = 0$ at the hole doping 0.125, then $d$-form factor is found to have largest magnitude at all energies. Lastly, we note that the magnitude of the $s$-form factor is comparable to others (although it is never the strongest channel), whereas experiment finds it to be smaller than the others.

\begin{figure}
\includegraphics[width=1\columnwidth]{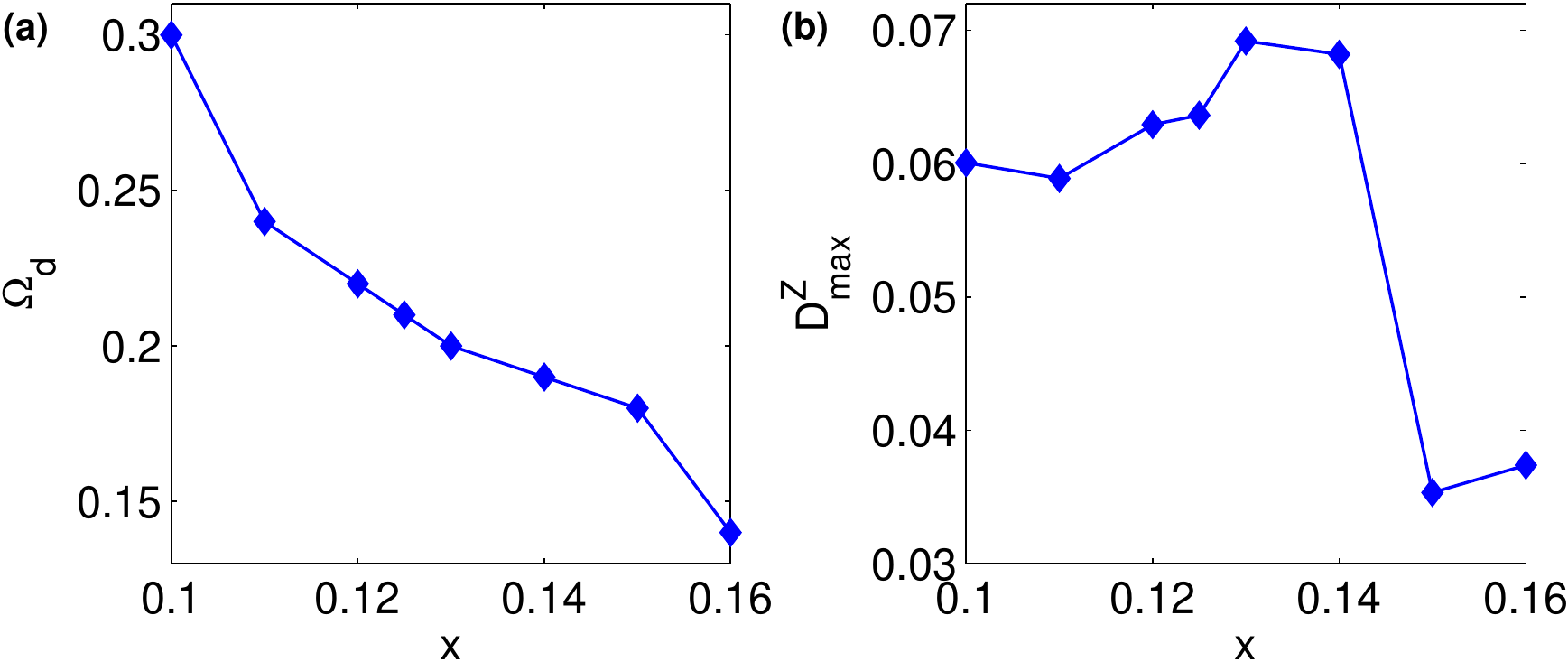}
\caption{ (Color online) Doping dependence of (a) energy at which $d$-form factor peaks ($\Omega_{d}$) and (b) corresponding magnitude ($D^{Z}_{max}$).}
\label{fig:nPDW_formFactors_doping}
\end{figure}

The doping dependence of the peak value of the $d$-form factor ($D^{Z}_{max}$) and corresponding bias ($\Omega_{d}$) is shown in Fig. \ref{fig:nPDW_formFactors_doping}. $\Omega_{d}$ decreases monotonically with hole doping. 
On the other hand, $D^{Z}_{max}$ shows a non-monotonic behavior as function of hole doping. First, it increases achieves a maximum at doping $x = 0.13$ and then drops rapidly. This is in agreement with the doping dependence of the STM intensity at the density wave modulation wave vector which can be thought as a measure of $d$-form factor magnitude \cite{Fujita2}.

\begin{figure}
\includegraphics[width=1\columnwidth]{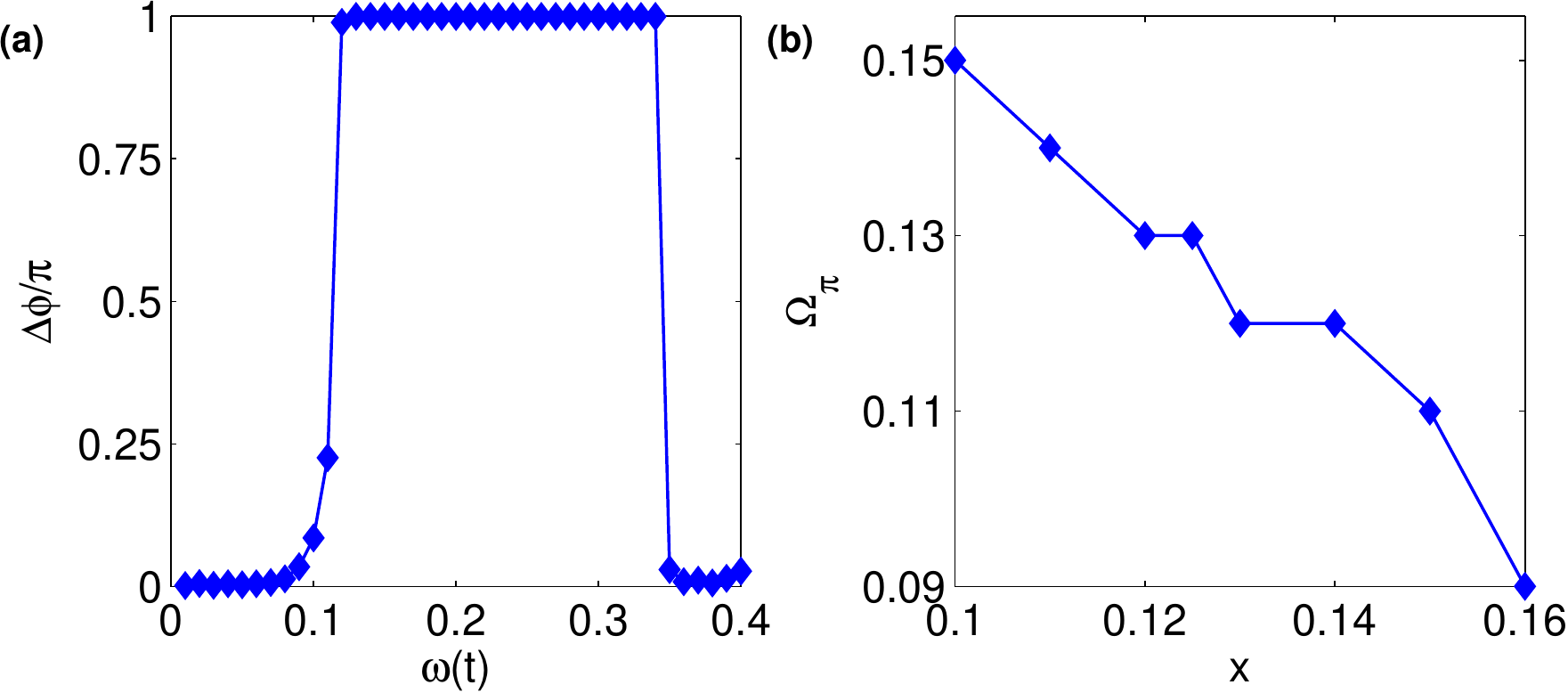}
\caption{ (Color online) (a) Bias dependence of average spatial phase difference defined in Eq. \ref{Eq:phase}. (b) Bias $\Omega_\pi$ at which initial $\pi$ phase jump in $\Delta \phi$ takes place vs. doping.    }
\label{fig:nPDW_phase_difference}
\end{figure}

The average spatial phase difference ($\Delta\phi$) between the $d$-form factor density wave modulations at positive and negative biases, computed using Eq. \ref{Eq:phase}, is shown in Fig. \ref{fig:nPDW_phase_difference}(a). We find at $x=0.125$ that in the vicinity of Fermi level spatial phase difference is zero and turns to $\pi$ for $ \omega  > 0.12t$. This bias dependence is in excellent agreement with the STM experiment \cite{Hamidian}. Fig. \ref{fig:nPDW_phase_difference}(b) shows that the energy ($\Omega_{\pi}$) at which $\pi$ phase shift occur decreases with hole doping. We note that in the supplementary information of Ref. 12, the authors show the bias dependence of $\Delta \phi$ at few more doping levels from which one can infer that the energy corresponding to $\pi$ phase shift decreases with increasing hole doping level, similar to what we observe for the nPDW state.

\begin{figure}
\includegraphics[width=1\columnwidth]{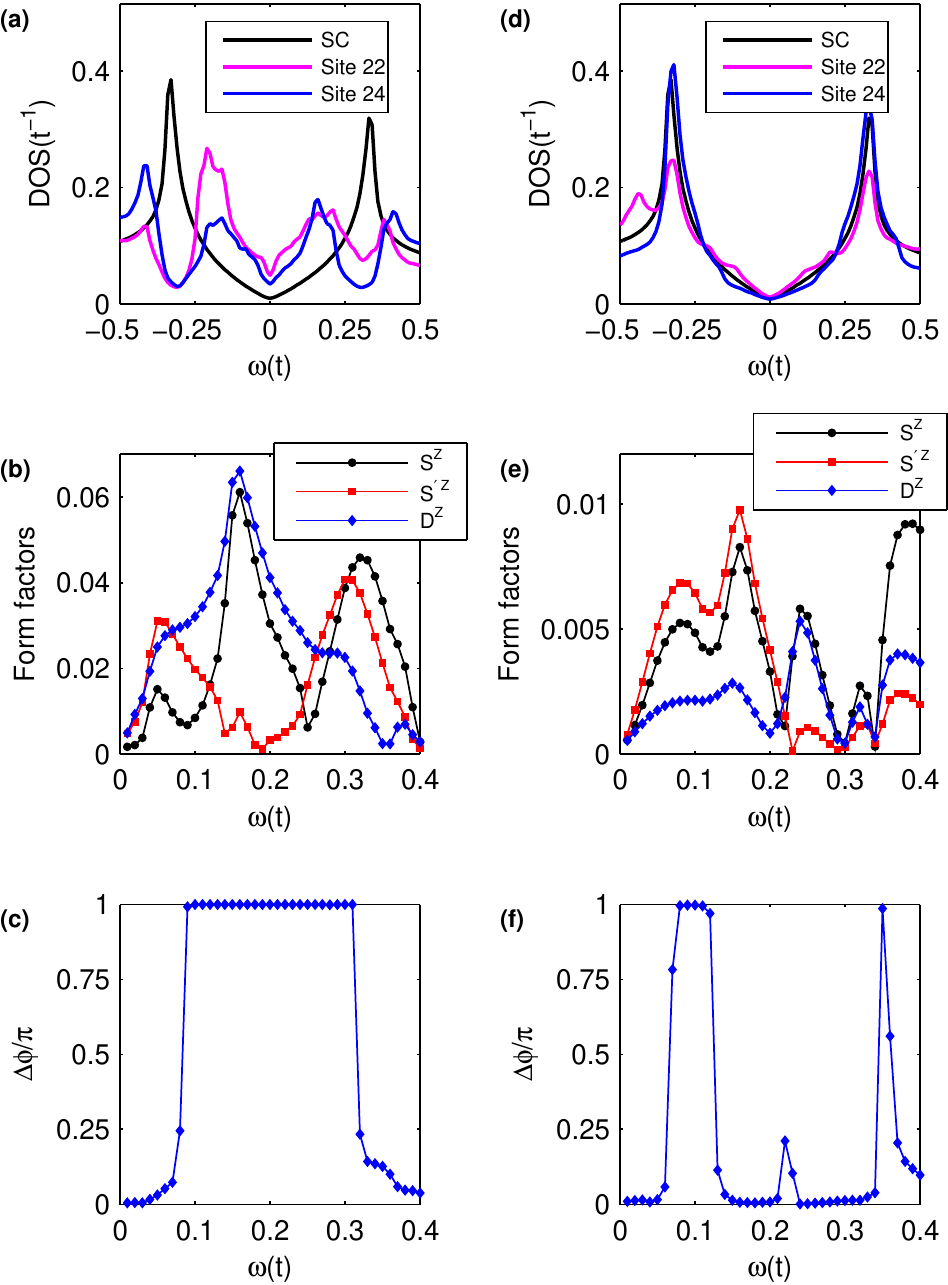}
\caption{ (Color online) (a)-(c) Lattice DOS, form factors and average spatial phase difference ($\Delta \phi$) in the case when nPDW charge and bond modulations are turned off keeping only pair field modulations. (d)-(f) Lattice DOS, form factors and average spatial phase difference ($\Delta \phi$), respectively, in the case when nPDW pair field modulations are turned off keeping charge and bond modulations. }
\label{fig:selective_modulations}
\end{figure}

To get a better understanding of the bias dependence of form factors and spatial phase difference, we attempt to disentangle PDW and CDW orders intertwined in the nPDW state, "by hand". We start with the self-consistent mean fields in the nPDW state discussed previously. As a first test, we do the following replacements in Eq. \ref{Eq:BdG}: $\delta_{i} \rightarrow \delta_{0}$ and $\chi^{v}_{i j \sigma} \rightarrow \chi^{v}_{0}$, where, subscript 0 indicates that the mean fields correspond to the uniform superconducting state. The pair field remains inhomogeneous and unchanged from the nPDW solution. In the second test, we do the following replacements in Eq. \ref{Eq:BdG}: $\Delta^{v}_{i j \sigma} \rightarrow \Delta^{v}_{0}$, and leave bond field and hole density inhomogeneous and unchanged from the nPDW solution.
The chemical potential is adjusted in both tests to yield the correct average electron filling. Results for the lattice LDOS, form factors and spatial phase difference in first and second tests are shown in Fig. \ref{fig:selective_modulations}(a)-(c), and (d)-(f) respectively. Comparing Fig. \ref{fig:nPDW_details}(c) with Fig. \ref{fig:selective_modulations}(a) and (d), we find that the two sets of coherence peaks in the nPDW state lattice LDOS are indeed originating from the pair density wave, and that the CDW has an insignificant effect. Fig. \ref{fig:selective_modulations}(b) shows that the $d$-form factor in the pure pair density wave state has the highest magnitude at the energy corresponding to one of the coherence peaks in the lattice LDOS ($\Omega_{d} = 0.16t$) and its bias dependence and overall scale is very similar to the nPDW state (Fig. \ref{fig:nPDW_formFactors}). However, when PDW order is artificially set to zero then $d$-form factor acquires a bias dependence and scale which is very different from the nPDW state as evident from Fig. \ref{fig:selective_modulations}(e). The importance of PDW is again manifested in Fig. \ref{fig:selective_modulations}(c) which shows that setting the charge density modulations to zero artificially has little effect on the spatial phase difference observed in the nPDW state (Fig. \ref{fig:nPDW_phase_difference}). However, when the PDW order is set to zero then we get a very different bias dependence of spatial phase difference as evident from Fig. \ref{fig:selective_modulations}(f). Thus we conclude that the most significant features in the bias dependence of lattice LDOS, $d$-form factor and spatial phase difference in the nPDW state are originating from the pair field modulations.

\section{Discussion}
\label{Discussion}
Within the inhomogeneous Gutzwiller approximation, for the parameters employed here,  the uniform $d$-wave superconducting state  has a lower energy than the charge ordered states at all doping levels. Thus the nPDW is not the ground state of the $t-t'-J$ model.
However, the energy difference between the uniform state and charge ordered states is of the order of only ${\cal O}$(1 meV) per site \cite{WeiLin,Yang}. Thus, it  is entirely plausible that other effects not included in the model such as disorder and electron-phonon interactions may stabilize these fluctuating charge ordered states \cite{Kivelson,Chou}. In fact, the short-ranged nature of these states, observed in STM \cite{Yazdani} and resonant elastic x-ray scattering experiments \cite{Damascelli}, suggests that disorder might be playing an important role. Different local disorder environments may then also pin slightly different states, resulting in slightly different local LDOS patterns that can be identified in STM images, not just two different ladder-type domains, as is normally assumed.

 As pointed out in Ref. 11, the evolution of the GW factors with doping  is responsible for the remarkable degeneracy of the various charge states shown in Figure \ref{fig:Eg_vs_doping} across the doping range. The energy splitting of these states above the homogeneous superconducting state remains almost the same across this range as well.  Thus the addition of a magnetic field on the order of 10T or 1meV per site can potentially stabilize long-range charge order. It is tempting to conclude that the recent observation of charge order in YBCO with a large correlation length, at a magnetic field of order 30T may be reflecting this effect \cite{Gerber, Jang}.

We find that at a given doping, nPDW states with different ordering wave vectors ${\bf Q}$ around $[0.3, 0]$ exist. Keeping the same initial guess and changing the system size ($N\times N$) results in charge ordered states with slightly different ${\bf Q} = [Q, 0],$ since Q is a multiple of $1/N$. However, LDOS, form factor and spatial phase difference results are insensitive to such small changes. All such states at nearby ${\bf Q}$ are extremely close in energy, and hence, at the level of the Gutzwiller approximation, we can not quantitatively address the doping dependence of the charge order wave vector. However, the bias dependence of the form factors and spatial phase difference is robust with respect to the change of ordering wave vector, band structure ($t'$) and doping. We always find a dominant $d$-form factor at higher energies and a shift of $\pi$ in the average spatial phase difference beyond a particular energy scale.

The analysis presented in the the previous section, whereby PDW and CDW order were artificially suppressed independently, strongly suggests that PDW character is necessary to explain the spectral characteristics, in particular the bias dependence  of the intra unit cell form factors and spatial phase difference in  experimental measurements on BSCCO.  

It is important to note further that the bias dependence of the form factors in the current theory is the clear result of electronic correlations in the CuO$_2$ plane.  It has been observed in x-ray spectroscopy that the plane in YBCO, for example,  buckles in a pattern of O displacements that mimics a $d$-wave form factor\cite{Hayden}, and suggested that this structural pattern imprints itself on the local tunneling conductance.  However, it is difficult to see how such a structural effect should be sensitive to the applied bias, as seen in experiment and predicted here.  Nor is it clear why, in such a scenario, the other form factors can be stabilized  in other bias ranges. 

Very recently, a new paper \cite{Mesaros} appeared analyzing the conductance maps in BSCCO according to a new type of ``phase resolved electronic structure visualization", concluding that the charge states observed were locally commensurate with lattice constant 4$a_0$ separated by phase slips (discommensurate), rather than incommensurate. The authors also stated that their findings were consistent with strong-coupling $r$-space based theories rather than Fermi surface driven instabilities. The  latter conclusion is consistent with a $t-J$ model description of the underdoped cuprates, and is quite consistent with ours. Since we have not included disorder or allowed for discommensuration, we cannot address their findings directly in this work.  However it seems intuitive that the introduction of disorder may favor discommensuration.  We leave this for a later project.


\section{Conclusions}
In summary, we have shown that there exist low-energy, commensurate and incommensurate charge modulated renormalized mean field solutions of the $t-t'-J$ model that are not the ground state at any filling, but which are extremely close to the energy of the uniform superconducting state.  Furthermore, the incommensurate charge ordered states, called nPDW 
is intertwined with modulated superconductivity, and display properties remarkably similar to STM observations of the 1D  modulated states seen on the surface of BSCCO and NaCCOC.   These are well-established features of cuprate physics that have intrigued workers in the field for almost a decade, but until now have defied explanation. Among these properties are the same spectra and pattern of tunneling conductance maps within the unit cell as observed by STM on under- to optimally doped BSCCO and NaCCOC. To calculate these patterns, as well as continuum LDOS spectra within the unit cell, we employed the new Wannier function-based method of Ref. \onlinecite{Choubey}, which enables the calculation of the wavefunctions in the correlated state at any 3D position, including several \AA ~ above the surface where the STM tip is placed. This gives us an unprecedented ability to compare with details of the experiments in the charge ordered regime.

In addition, the bias dependence of intra-unit cell $d$-, $s'$- and $s$-form factors and their spatial phase difference  were obtained in the nPDW state and display good agreement with the STM observations.  The energy of the peak $d$-wave form factor depends with doping in a manner similar to the pseudogap. Note that with the exception of Ref. \onlinecite{WeiLin}, previous theories of charge ordered states in $t-t'-J$ type models treated only commensurate (4 unit cell wavelength) charge order states, and could express observables only in terms of bias-independent bond variables.

We have furthermore argued that, while the incommensurate states found here are not the ground state of the system studied, relatively small perturbations can stabilize them.  In particular, we discussed the  possibility that impurities stabilize the charge order, leading to the disordered 1D patterns observed in STM on BSCCO and NaCCOC. This disordered ground state would also be consistent with the short-range charge-order observed by RIXS. In such a system, a magnetic field should suppress superconductivity and eventually favor long-range charge order, as observed in experiments. We leave investigation of this intriguing scenario to a future study.

\vskip .5cm
{\it Acknowledgements.}  
The authors are grateful for useful discussions
with {B. M. Andersen, J. C. Davis, K. Fujita, S. Hayden, A. Kreisel, S. Maiti, C. P\'{e}pin, T.M. Rice, S. Sachdev, A.-M. S. Tremblay and S. Verret.}    {PC and PJH were supported by NSF grant NSF-DMR-1407502. PJH's work was  performed  in part at the Aspen Center for Physics, which is supported by National Science Foundation grant PHY-1066293. WLT and TKL were supported by Taiwan Ministry of Science and Technology Grant 104-2112-M-001-005. Part of calculation was supported by the National Center for High Performance Computing in Taiwan.

\end{document}